# Electronic nature of coverage-dependent nanosurface effect by cooperative orbital redistribution

Guolei Xiang,[1†] Yan Tang,[2] Zigeng Liu,[3†] Wei Zhu,[2] Haitao Liu,[4] Jiaou Wang,[5] Guiming Zhong,[6] Jun Li,[2] Xun Wang[2†]

1. School of Science, Beijing University of Chemical Technology, Beijing 100029, P.R. China.
2. Department of Chemistry, Tsinghua University, Beijing 100084, P.R. China.
3. Max-Planck-Institute for Chemical Energy Conversion, Stiftstrasse 34-36, D-45470 Mülheim an der Ruhr, Germany.
4. Department of Chemistry, University of Pittsburgh, Pittsburgh, Pennsylvania 15260, United States.
5. Beijing Synchrotron Radiation Facility, Institute of High Energy Physics, Chinese Academy of Science, Beijing 100049, P.R. China.
6. Xiamen Institute of Rare Earth Materials, Chinese Academy of Sciences, Xiamen 361021, Fujian, P.R. China.

†Corresponding email:
   xianggl@mail.buct.edu.cn; zigeng.liu@cec.mpg.de; wangxun@tsinghua.edu.cn.

**Abstract:**
Nanomaterial surface states can effectively modify or even dominate their physical and chemical properties due to large surface-to-volume ratios. Such surface effects are highly dependent on particle size and ligand coverage, yet the underlying electronic-level mechanism still remains unknown. Using $TiO_2$ nanosheet as a model system, we reveal the electronic nature of coverage-dependent nanosurface effects through varying ligand coverage and probing the modified surface bonding and electronic band structures with near-edge X-ray absorption fine structure. We discover experimentally that surface ligands can competitively polarize the $3d$ orbitals of surface Ti atoms into chemisorption states, which is cooperative with increased ligand coverages. Such coverage-dependent cooperative orbital redistribution accounts for various nanosurface effects on regulating the electronic structure, surface reactivity, optical property, and chemisorption of nanomaterials.

Chemisorption is crucial for catalysis, electrochemistry, crystal growth, surface electron transfer, etc. At the nanometer scale, particle size and ligand coverage can significantly affect the surface chemistry of nanomaterials apart from the chemical structures of solid surfaces and ligands (*1, 2*). In particular, the surface ligation states can even dominate physical and chemical performances of nanomaterials owing to the extremely large surface-to-volume ratios (*1, 3-5*). Typically, sub-5-nm nanomaterials can often display size-dependent catalytic capability (*6, 7*). The optical properties of semiconductor nanoparticles (NPs) and metal nanoclusters (NCs) can also be influenced by both particle size and ligand coverage (*1, 8-14*). For example, the photoluminescent quantum yield (PLQY) of CdSe NPs can be enhanced by increasing the coverages of oleic acids (*9*); the photoluminescent wavelength of thiolated Au NCs can be modified by varying ligands coverages (*8*). The physical nature driving the origin of these nano effects in the surface science of nanomaterials, however, has long been enigmatic, because they cannot be simply explained by the geometrical parameters of decreased size, increased specific surface area or defect ratios (*1, 15-17*).

Fundamentally, the nano effects regarding nanosurface chemistry ultimately originate from the varied electronic structures through surface coordination bonds with the ligands (*16, 18, 19*). Thus a general understanding of the mysteries in nanosurface chemistry (surface reactivity, size effect, and coverage effect) must rely on elucidating the electronic-level mechanisms of molecule-surface interactions at the nanometer scale. But the electronic-level insights into nanosurface effects are experimentally hindered by the difficulties in selectively probing surface ligation bonds and the altered electronic structures of nanomaterials. This goal relies on both characterization technologies and ideal material models because chemisorption can only effectively perturb the bonding and electronic structures of top-layer atoms, and the interfering signals from the bulk phase should be highly reduced (*20, 21*). Although scanning tunneling microscopy (STM) and photoelectron spectroscopies (PES) can study the physicochemistry on flat surfaces of single crystals (*22, 23*), it is challenging to effectively probe the surface chemical bonds and perturbed surface electronic states of most nanomaterials.

Two-dimensional (2D) nanosheets (NSs) with atomic thicknesses are rising as promising models to study the mechanisms of surface chemistry and catalysis due to the definite surface structures (*24, 25*). With almost all the atoms locating on the surfaces, 2D NSs are ideal systems to investigate the electronic mechanisms of ligation effects and coverage effects in nanosurface science, as the interfering signals from the bulk can be highly decreased. Using 0.4-nm-thick $TiO_2$(B) NSs as a model system, herein we illuminate how ligand coverage tunes the optical absorption and electronic structures of nanomaterials. Through characterizing the varied surface bonding and electronic structures by near edge X-ray absorption fine structure (NEXAFS) spectroscopy, we show that surface ligands can effectively confine the $3d$ orbitals of surface Ti atoms into chemisorption states. This interaction competitively redistributes the surface atomic orbitals from delocalized energy band states to localized chemisorptive bonds, which is cooperative and coverage-dependent on the nanoscale. We propose that increasing ligand coverage can modify the optical properties of surface states, narrow electronic energy bands, enhance the surface reactivity and strengthen chemisorption on nanomaterials.

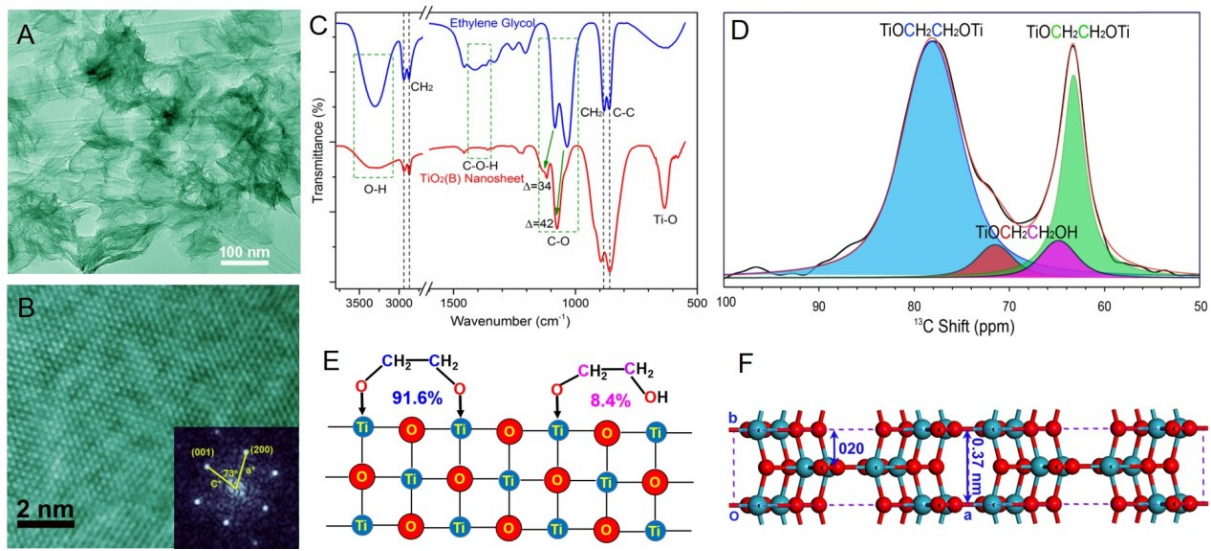

**Fig. 1. Structural model and surface chemistry of TiO₂(B) nanosheets capped with ethylene glycol (EG).** (**A**) TEM image of TiO₂(B) nanosheets. (**B**) HRTEM and FFT pattern to index the exposed (010) facet. (**C**) FTIR spectra showing the dissociative bonding feature of EG ligands. (**D**) Fitted $^{13}$C ssNMR spectrum to calculate the ratios of bidentate to monodentate EG ligands. (**E**) Scheme of the bidentate and monodentate adsorption configurations of EG and the corresponding coverages. (**F**) Atomic structure model of TiO₂(B) nanosheets with a thickness of ~0.37 nm in the *b* dimension.

TiO₂(B) nanosheets (Fig. 1A and fig. S1) were prepared by hydrolyzing TiCl₃ in ethylene glycol (EG) (*26*) and expose (010) facet as determined by indexing the fast Fourier transformation (FFT) of a high-resolution transmission electron microscopy image (HRTEM, Fig. 1B and fig. S2). Fourier transform infrared spectra (FTIR, Fig. 1C and fig. S3) of EG on the NSs show that the O-H modes at 3300 cm⁻¹ and 1400 cm⁻¹ are highly reduced and C-O modes shift towards higher wavenumbers by 42 and 34 cm⁻¹, respectively, while the positions of C-C and CH₂ modes remain unchanged. The results indicate that EG molecules bind to surface Ti sites by dissociating -OH groups, which is further confirmed by $^{13}$C solid-state nuclear magnetic resonance (ssNMR) spectra (Fig. 1D). The ssNMR results show that the $^{13}$C resonance arising from CH₂ at 63.2 ppm in EG shifts to lower fields at 78.0 ppm in NSs (*27*). By comparing the deconvolution results (Fig. 1D) we assign the 78.1-ppm resonance to bidentate EG ligands, the resonance at 64.8 ppm and the shoulder peak at 71.5 ppm to HO-CH₂- and -CH₂-O-Ti of monodentate EG ligands, and the resonance at 63.2 ppm to physisorptive EG molecules, respectively. Our calculations show that about 91.6% surface Ti sites are occupied by bidentate EG based on the integrated intensity ratio of bidentate to monodentate ligands (Fig. 1E and fig. S4). Based on the ssNMR EG coverages and saturated adsorption models, we further calculated the weight ratios of EG to TiO₂ for TiO₂(B) NSs with varied thicknesses in the *b* dimension (fig. S5 and table S1). The $w$(EG):$w$(TiO₂) ratio of TiO₂(B) NSs is ~27.3% as characterized by thermal gravimetric analysis (TGA, fig. S6), which matches the calculated value of 27.2% for a 3-layer model (table S1). Therefore, the TiO$_2$(B) NSs are composed of 3 [TiO] atomic layers in the *b* dimension (0.37 nm, Fig. 1F).

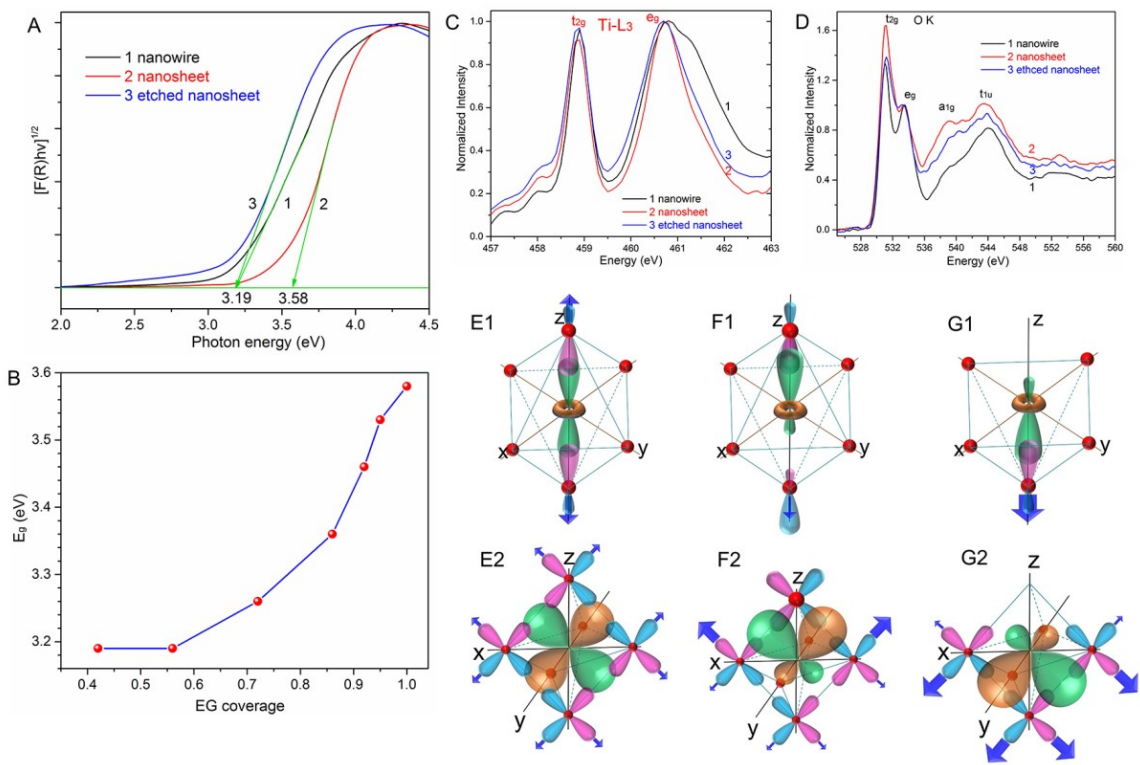

**Fig. 2. Coverage effects of EG ligands on the band gaps of TiO$_2$(B) NSs and the electronic mechanisms.** (**A**) Band gaps of TiO$_2$(B) nanowires (3.19 eV), nanosheets (3.58 eV) and etched nanosheets (3.19 eV). (**B**) Dependence of nanosheet band gaps upon EG coverages. NEXAFS data of Ti-L$_3$ (**C**) and O-K (**D**) edges normalized to the pre-edges and $e_g$ peaks (see below). (**E-G**) Distribution configurations of surface Ti 3*d* orbitals in octahedral ligand fields with different top ligands in (**E**) nanowires, (**F**) nanosheets and (**G**) etched nanosheets. (**E1-G1**) show varied bonding features of σ-type 3$d_{z^2}$; (**E2-G2**) show varied bonding features of π-type 3$d_{xz}$ and 3$d_{yz}$. The arrows indicate the relative delocalization tendencies of 3*d* orbitals.

Optical absorption, characterized by UV-visible diffuse reflectance spectroscopy (DRS) and further transformed with Kubelka–Munk function (Fig. 2A and fig. S7) (*28*), was used to investigate coverage effects on modifying nanomaterial electronic structures. A series of TiO$_2$(B) nanosheet samples with varied EG coverages (*θ*) were prepared by etching off the ligands with HNO$_3$, and their shapes, long-range ordered crystal structures and local atomic configurations keep unchanged as characterized by TEM, X-ray diffraction (XRD) and extended X-ray absorption fine structure (EXAFS) (fig. S8, S9). TiO$_2$(B) nanowires (NWs, fig. S10) with diameters between 50 and 200 nm were chosen as a bulk control. The electronic band gap of TiO$_2$(B) NWs is 3.19 eV (Fig. 2A), a typical value of TiO$_2$, and it increases to 3.58 eV for raw TiO$_2$(B) NSs saturated by EG ligands and decreases to 3.19 eV after etching off 58% EG (Fig. 2A). This phenomenon suggests that the band gaps of TiO$_2$(B) NSs are affected by EG coverages. We therefore performed more detailed studies by combining optical absorption and coverage quantification of EG through quantifying the amounts of C in the NSs (fig. S11 and table S2). As shown in Fig. 2B, the band gaps vary from 3.19 eV when $\theta < 0.56$ to 3.58 eV when $\theta = 1$. At low coverages ($\theta < 0.56$), etched TiO$_2$(B) NSs show bulk-like band gaps, indicating the ligation interactions are not strong enough to modify the bulk electronic structures. However, this surface effect can be enhanced effectively by increasing EG coverages, especially when $\theta > 0.90$. At high coverages, the band gaps are more sensitive to EG coverages, suggesting the surface ligation states dominate the total electronic structures and optical properties.

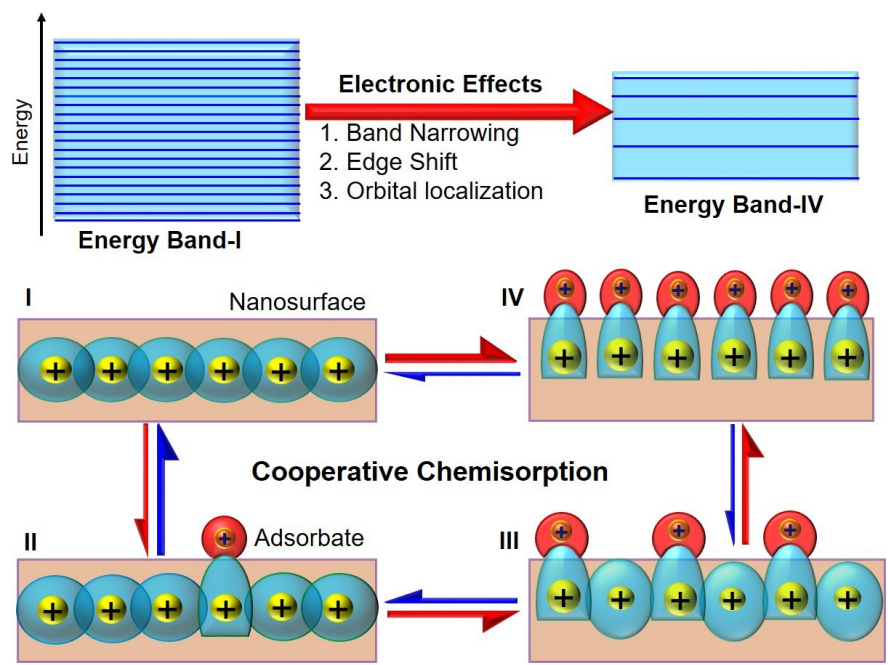

**Fig. 3. Scheme illustration of the electronic nature of coverage-dependent surface effects through cooperative chemisorption.** Without adsorbates (I), nanomaterials show intrinsic electronic structures, in which surface atomic orbitals (SAOs) preferentially delocalize into the lattices. Adsorbates can induce SAO redistributions by increasing ligand coverages (I→II→III→IV). At high coverages (IV), SAOs are highly pinned in surface chemisorption states. The cooperative orbital redistributions can narrow the energy bands, shift band edges, and confine SAOs in chemisorption states. Further, both surface reactivity and chemisorption strength will be enhanced by increasing coverages.

To reveal the underlying electronic mechanism of such coverage-dependent surface effects, we characterized the varied bonding and electronic structures with NEXAFS (Fig. 2C, 2D and fig. S12) to probe the densities of unoccupied electronic states. For TiO$_2$, the empty states arise from antibonding orbitals formed by Ti 3$d$, 4$s$, 4$p$ with O 2$p$ atomic orbitals (AOs), which locate in the conduction band and further split into four sub-bands of $t_{2g}$, $e_g$, $a_{1g}$ and $t_{1u}$ symmetries at the one-electron level. In an octahedral ligand field, the five-fold degenerate Ti 3$d$ AOs split into two groups of molecular orbitals (MOs): $\sigma$-type 3$d_\sigma$ orbitals in $e_g$ sub-band evolved from $d_{x^2-y^2}$ and $d_{z^2}$ AOs and $\pi$-type 3$d_\pi$ orbitals in $t_{2g}$ sub-band from $d_{xy}$, $d_{xz}$ and $d_{yz}$ orbitals (fig. S13) (*29*). Ti-L and O-K edges correspond to electronic dipole transitions of Ti 2$p$→Ti 3$d$ and O 1$s$→O 2$p$, respectively Changes in the peak width of $e_g$ band and the relative intensity ratio of $t_{2g}$ to $e_g$ ($It_{2g}/Ie_g$) directly echo the varied local and extended bonding features of Ti 3$d$ AOs and the ratios of Ti 3$d$ and O 2$p$ in unoccupied electronic states (*30*).

For TiO$_2$(B) nanowires, the Ti-L$_3$ $e_g$ peak splits due to the geometry distortion of [TiO$_6$] octahedron (Fig. 2C and fig. S12); the $e_g$ band is wider than $t_{2g}$ band because $\sigma$ bonds are more overlapped and delocalized (Fig. 2E1) than $\pi$ bonds (Fig. 2E2); the $It_{2g}/Ie_g$ ratio of nanowire O-K edge (3:2, Fig. 2D) is consistent with the results of bulk TiO$_2$ (*31*). Therefore, TiO$_2$(B) nanowires can be taken as a bulk model, whose $\sigma$-type 3$d_\sigma$ orbitals ($d_{z^2}$) in $e_g$ band and $\pi$-type 3$d_\pi$ orbitals ($d_{xz}$ and $d_{yz}$) in $t_{2g}$ band distribute uniformly in [TiO$_6$] octahedral ligand fields (Fig. 2E).

The electronic features of coverage effects are revealed by comparing the NEXAFS results of nanowires, nanosheets and etched nanosheets (Fig. 2C, 2D, fig. S12). Compared to nanowires, the Ti-L$_3$ $e_g$ peaks of nanosheets and etched nanosheets obviously become narrower and the splitting decreases, indicating reduced overlaps and delocalization of $d_{z^2}$ orbitals in TiO$_2$(B) lattices. The narrowed $e_g$ bands mainly result from size reduction, because the atomic thickness removes long-range lattice periodicity and effectively suppresses the delocalization of $d_{z^2}$ orbitals to form $e_g$ bands. The Ti-L$_3$ $e_g$ peak of etched nanosheets is broader than that of raw nanosheets (Fig. 2C), so surface ligands must also contribute to the suppressed delocalization of $d_{z^2}$ orbitals and narrowed $e_g$ bands. For nanosheets, the 3$d_{z^2}$ orbitals of surface Ti atoms are partly polarized and localized into chemisorptive states with EG ligands, thus the delocalization is further decreased (Fig. 2F1). For etched nanosheets, most five-fold coordinated surface Ti atoms (Ti$_{5f}$) locate in pyramidal configurations, thus their 3$d_{z^2}$ orbitals are less polarized into chemisorption states but tend to extend into the lattice to enhance and broaden $e_g$ bands (Fig. 2G1) (*30*).

Coverage effects on $\pi$ bonds are revealed by analyzing the varied $It_{2g}/Ie_g$ of Ti-L$_3$ and O-K edges. Compared with nanowires, the Ti-L$_3$ $It_{2g}/Ie_g$ decreases from 0.96 to 0.90 for nanosheets, suggesting reduced ratios of unoccupied Ti 3$d_\pi$ states in $t_{2g}$ bands, while the increased O-K $It_{2g}/Ie_g$ from 1.32 to 1.64 suggests increased ratios of unoccupied O 2$p_\pi$ states in $t_{2g}$ bands. The difference further indicates that EG ligands also bond to Ti atoms through overlapping their O 2$p_\pi$ orbitals with Ti 3$d_\pi$ orbitals, in which O 2$p_\pi$ shares electron pairs with Ti 3$d_\pi$. As a result, Ti 3$d_\pi$ orbitals also polarize into chemisorption states, which decreases their delocalization into the lattices to form $t_{2g}$ bands (Fig. 2F2). For etched nanosheets, $It_{2g}/Ie_g$ of Ti-L$_3$ and O-K are close to that of nanowires, suggesting they possess similar $t_{2g}$ bands, which matches their equal band gaps (Fig. 2A). It follows that Ti 3$d_\pi$ AOs also tend to extend into the lattices to form $t_{2g}$ bands without confinement of EG ligands (Fig. 2G2).

The above NEXAFS results show that surface EG ligands can competitively redistribute surface Ti 3$d$ orbitals of TiO$_2$(B) nanosheets through polarizing them from delocalized energy band states to localized chemisorptive states. To understand the general electronic features of nanosurface chemistry, we introduce a parameter, the distribution fraction (*f*) of a surface atomic orbital (SAO), to further reveal such competitive orbital redistributions (Fig. 3). For surface atoms, their valence SAOs can mainly distribute into two parts, confined into localized surface states ($f_S$) or delocalized into extended systems to form Bloch states in energy bands ($f_B$). Since AOs are normalized, we have $f_S+f_B=1$. Assuming the confined surface states correspond to chemisorption, then $f_S$ is a descriptor measuring surface reactivity and the strength of chemisorption, while $f_B$ represents the contribution to the energy band and correlates to band width.

The quantum normalization feature of AOs implies $f_S$ and $f_B$ are competitive, and cooperativity among adsorbates can exist to simultaneously redistribute SAOs at different surface sites. For adsorbates, such cooperative effects are coverage-dependent as shown in Fig. 3. On clean surfaces (I) SAOs mainly delocalize into the lattice, thus nanomaterials show their intrinsic electronic structures. At low coverages (II), adsorbates partly polarize SAOs into chemisorption states, but cannot effectively modify band structures. The perturbation effects can be enhanced by increasing adsorbate coverages (II→III), which confines more SAO fractions in surface states and decreases their delocalization through increasing $f_S$ while reducing $f_B$. For nanostructures with extremely large surface-to-volume ratios, such as $TiO_2$(B) nanosheets, both $f_S$ and $f_B$ reach the extrema at highly saturated coverages (IV), then adsorbates can effectively extract SAOs from energy bands to form strong chemisorption interactions. We propose that the electronic nature of coverage-dependent surface effects lies in cooperatively increasing $f_S$ but competitively decreasing $f_B$ upon increasing ligand coverage.

Such coverage-dependent competitive and cooperative orbital redistributions can cause two basic effects. For nanomaterials, delocalization of SAOs into the lattices gets decreased, which can narrow the energy bands, shift band edges, modify the electronic structures and increase surface reactivity (Fig. 3). Specifically, the band gaps and optical properties of semiconductors can be tuned as we have shown in Fig. 2A and 2B. For adsorbates, the strength of chemisorption can be enhanced due to the increased $f_S$, and the increased surface reactivity can further accelerate more chemisorption. In brief, increasing adsorbate coverages can enhance both surface reactivity and chemisorption strength, which is a positive feedback and can eventually lead to saturated surface ligation. We refer to this feature of coverage-dependent molecule-surface interactions on nanomaterials as nanoscale cooperative chemisorption.

We further study the coverage effects through quantum chemical calculations based on density functional theory (DFT) to analyze the tunable bonding interactions. The adsorption structures with different EG coverages ($\theta$ = 25%, 50%, 75% and 100%) on $TiO_2$(B) NSs were optimized (fig. S14), and the model of 25% coverage was selected to analyze the bonding configurations between surface Ti atoms and the O atoms in EG ligand using periodic natural bond orbital (NBO) analysis. As shown in table S3, EG molecules bind to Ti sites through triple bonds, one $\sigma$ and two $\pi$ bonds. The $\sigma$ bonding dominates the interaction because its orbital hybridization 0.13 Ti($s^{0.5}p^{1.18}d$) + 0.87 O($sp^{0.72}$) is apparently greater than the $\pi$ bonds of 0.08 Ti($s^{0.06}p^{0.508}d$) + 0.92 O($s^{0.02}p$) or 0.06 Ti($s^{0.11}p^{0.91}d$) + 0.94 O(p). The bonding energies between Ti sites and O atoms in EG are used to evaluate the strength of coverage-dependent chemisorption (Fig. 4 and table S4). We find that the Ti-O bonding energy increases with increased EG coverages, further supporting that interaction between EG and TiO$_2$(B) NSs can be enhanced by increasing EG coverages. The increased bonding energies indicate chemisorption at nanometer scale can be enhanced by increasing ligand coverages. This scenario of metal-ligand surface bonding is in line with the trend of chemisorption of CO on heterogeneous single atom catalysts with metal-oxide support (*32*), and coverage-dependent surface reactivity towards ligand exchange of Ag NCs (*33*), indicating it is a general occurrence in nanosurface science.

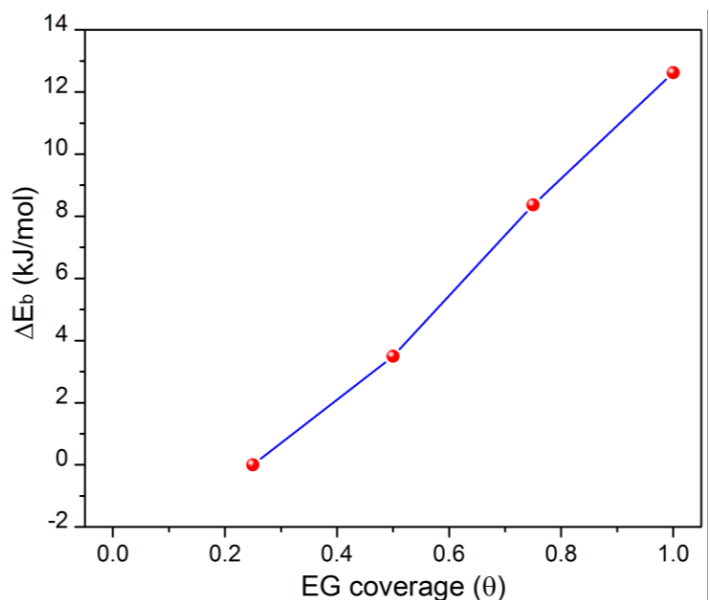

**Fig 4.** Calculated average Ti-O bonding energies (E$_b$) as a function of EG coverages. The zero point is set at the coverage of 0.25.

This competitive and cooperative orbital redistribution by chemisorption is clearly a general surface phenomenon that has been underappreciated in the nanomaterial and catalysis community. The interplay between the localized orbitals in the adsorptive states and the delocalized orbitals in the band structure deems to play a significant role in surface science and chemisorption. Our current finding of the electronic origin that drives coverage-dependent surface effects on nanomaterials provides a new avenue to tune electronic structures of nanomaterials and design of catalysts. The physical picture of cooperative chemisorption and the electronic-level mechanism of cooperative orbital redistribution present a viewpoint to reveal the mechanisms of nanoscale phenomena regarding molecule-surface interactions.

**ACKNOWLEDGMENTS**
We thank Hongyu Sun for taking TEM images, Yang Yang for measuring FTIR, and Chong Guo for collecting UV-vis spectra. We acknowledge the helpful discussions, suggestions and supports from Haifeng Dong, Yanggang Wang, Yong Long, O. Scherman, J. Baumberg, C. Grey, K. Griffith, Hua Wang, Lang Jiang, Xiangfeng Duan, Lirong Zheng, Xiaoming Sun and Dandan Han. This research was supported by Fundamental Research Funds for Central Universities of China (buctrc201812), National Natural Science Foundation of China (21431003, 21521091, 21590792, 91645203). The calculations were performed by using supercomputers at Tsinghua National Laboratory for Information Science and Technology and the Supercomputing Center of Computer Network Information Center of the Chinese Academy of Sciences. G. X. and X. W. initiated the research. G. X. conceived the ideas, designed the experiments, prepared the materials and wrote the manuscript. Z. L., W. Z., H. L. and X. W. helped with designing the experiments, analyzing the data and writing the manuscript; G. Z. carried out ssNMR measurement and elemental analysis; J. W. and G. X. carried out NEXAFS and EXAFS measurements and analyzed the results; Y. T. and J. L. carried out DFT calculations. All authors comment on the manuscript. The authors declare that there are no conflicts of interest.## Reference

1. M. A. Boles, D. Ling, T. Hyeon, D. V. Talapin, *Nat. Mater.* **15**, 141-153 (2016).
2. E. Roduner, *Chem. Soc. Rev.* **35**, 583-592 (2006).
3. J. Owen, *Science* **347**, 615-616 (2015).
4. P. R. Brown *et al.*, *ACS Nano* **8**, 5863-5872 (2014).
5. S. G. Kwon *et al.*, *Nano Lett.* **12**, 5382-5388 (2012).
6. M. Valden, X. Lai, D. W. Goodman, *Science* **281**, 1647-1650 (1998).
7. W. E. Kaden, T. P. Wu, W. A. Kunkel, S. L. Anderson, *Science* **326**, 826-829 (2009).
8. J. B. Liu *et al.*, *Angew. Chem. Int. Edit.* **55**, 8894-8898 (2016).
9. N. C. Anderson, M. P. Hendricks, J. J. Choi, J. S. Owen, *J. Am. Chem. Soc.* **135**, 18536-18548 (2013).
10. C. H. M. Chuang, P. R. Brown, V. Bulovic, M. G. Bawendi, *Nat. Mater.* **13**, 796-801 (2014).
11. B. P. Bloom *et al.*, *J. Phys. Chem. C* **117**, 22401-22411 (2013).
12. G. L. Wang *et al.*, *J Phys Chem B* **110**, 20282-20289 (2006).
13. C. Landes, M. Braun, C. Burda, M. A. El-Sayed, *Nano Lett.* **1**, 667-670 (2001).
14. Z. K. Wu, R. C. Jin, *Nano Lett.* **10**, 2568-2573 (2010).
15. M. Haruta, *Catal. Today* **36**, 153-166 (1997).
16. Y.-G. Wang *et al.*, *J. Am. Chem. Soc.* **138**, 10467-10476 (2016).
17. J. Kleis *et al.*, *Catal. Lett.* **141**, 1067-1071 (2011).
18. M. V. Ganduglia-Pirovano, A. Hofmann, J. Sauer, *Surf. Sci. Rep.* **62**, 219-270 (2007).
19. B. Hammer, J. K. Norskov, *Surf. Sci.* **343**, 211-220 (1995).
20. E. Shustorovich, *Surf. Sci.* **6**, 1-63 (1986).
21. B. Hammer, J. K. Norskov, *Adv. Catal.* **45**, 71-129 (2000).
22. O. Bikondoa *et al.*, *Nat. Mater.* **5**, 189-192 (2006).
23. B. Li *et al.*, *Science* **311**, 1436-1440 (2006).
24. Y. Sun, S. Gao, F. Lei, Y. Xie, *Chem. Soc. Rev.* **44**, 623-636 (2015).
25. D. H. Deng *et al.*, *Nat. Nanotechnol.* **11**, 218-230 (2016).
26. G. L. Xiang, T. Y. Li, J. Zhuang, X. Wang, *Chem. Commun.* **46**, 6801-6803 (2010).
9